\documentclass[prl,twocolumn,floatfix,showpacs ]{revtex4-1}
\UseRawInputEncoding
\usepackage{amsmath}
\usepackage{amssymb}
\usepackage{graphicx}% Include figure files
\usepackage{dcolumn}% Align table columns on the decimal point
\usepackage{bm}% bold math
\usepackage[colorlinks=true,linkcolor=blue,anchorcolor=blue, citecolor=cyan,urlcolor=cyan]{hyperref}% add hypertext capabilities
\usepackage[mathlines]{lineno}% Enable numbering of text and display math
\usepackage{ulem}
\usepackage{epstopdf}
\begin{document}
%========================================================
\title{Hidden altermagnetism}
\author{San-Dong Guo}
\email{sandongyuwang@163.com}
%\author{Ping Li$^{2}$ and  Guangzhao Wang$^{3}$}
\affiliation{School of Electronic Engineering, Xi'an University of Posts and Telecommunications, Xi'an 710121, China}
%\affiliation{$^2$State Key Laboratory for Mechanical Behavior of Materials, School of Materials Science and Engineering, Xi'an Jiaotong University, Xi'an, Shaanxi 710049, China}
%\affiliation{$^3$Key Laboratory of Extraordinary Bond Engineering and Advanced Materials Technology of Chongqing, School of Electronic Information Engineering, Yangtze Normal University, Chongqing 408100, China}
%\affiliation{ address 3}
%\date{\today}
%=======================Abstract===============================================================
\begin{abstract}
Hidden spin polarization (HSP) with zero net spin polarization
in total but non-zero local spin polarization has been proposed in certain nonmagnetic centrosymmetric compounds,  where the individual sectors forming the inversion partners are all inversion asymmetry. Here, we extend this idea to  antiferromagnetic materials with $PT$ symmetry (the joint symmetry of space inversion symmetry ($P$) and time-reversal symmetry ($T$)), producing zero net spin polarization in total, but either of the two inversion-partner sectors possesses altermagnetism, giving rise to non-zero local spin polarization in the real space, dubbed "hidden altermagnetism".
By first-principle
calculations, we predict that $PT$-symmetric bilayer $\mathrm{Cr_2SO}$ can  serve as a
possible candidate showing altermagnetic HSP. By applying an external electric field to break the global $P$ symmetry, the hidden altermagnetism can be separated and observed experimentally. Our works extend the hidden physics, and will also advance the theoretical and experimental search for new type of spin-polarized materials.

\end{abstract}
\maketitle
%\tableofcontents
%========================Introduction===========================================================
\textcolor[rgb]{0.00,0.00,1.00}{\textbf{Introduction.---}}
Spin-orbit coupling (SOC) can induce a momentum-dependent spin-splitting in noncentrosymmetric nonmagnetic materials, including  two conventional types: the Dresselhaus-type due to bulk inversion asymmetry  and the Rashba-type  in two-dimensional (2D) heterostructures due to structural inversion asymmetry\cite{h1,h2}.
 Although centrosymmetric  nonmagnetic systems are supposed to lack spin-splitting, there is a large
class of systems whose global crystal symmetry is indeed
centrosymmetric, but they consist of noncentrosymmetric  individual sectors,   producing  visible spin-splitting effects in the real space,   dubbed "hidden spin polarization (HSP)"\cite{h3}. Subsequently,  a number of
layered materials exhibiting HSP are predicted by the first-principles calculations\cite{h4,h5,h6}.
The HSP has been experimentally observed in many bulk materials\cite{h7,h8,h9,h10}, and  has also been reported in  monolayer $\mathrm{PtSe_2}$ with the spin-layer locking by the measurement of spin- and angle-resolved
photoemission spectroscopic\cite{h11}, which  triggers  more researches  on broader hidden physical effects\cite{h12}, such as hidden orbital polarization and hidden Berry curvature\cite{h121,h122}.
A natural question is whether one of the two inversion-partner sectors
 can possess other special properties that can also give rise to HSP. The answer lies in altermagnetism.

The altermagnetism exhibits alternating spin polarization with $d$-, $g$-, or $i$-wave symmetry  in Brillouin zone (BZ)\cite{k4,k5}.
With compensated antiparallel magnetic order, the
band structure of altermagnetism breaks time-reversal symmetry ($T$) and possesses   spin-splitting without the help of SOC\cite{k4,k5,k511,k512,k513}.
Although altermagnetism has  no net magnetization, unlike conventional
antiferromagnetism, the sublattices with opposite spins are connected by  rotational/mirror transformation
rather than by translation or inversion\cite{k4}.
 The concept of altermagnetism has also been extended to accommodate non-collinear spins and multiple local-structure variations\cite{k51}.
 The altermagnetism not only shares certain key properties with antiferromagnetism, but also it demonstrates even more similarities with ferromagnetism due to the alternating spin-splitting of the bands.
A number of altermagnetic materials  exhibiting momentum-dependent spin-splitting have
been revealed  both experimentally and theoretically\cite{h13}.
Recently, twisted altermagnetism, takeing one of all five 2D  Bravais lattices, has also been proposed in twisted magnetic Van der Waals (vdW) bilayers\cite{k8,k80}, and  a vertical electric field can induce valley polarization due to valley-layer coupling\cite{k9,k10}.
 Recently, a novel  antiferroelectric altermagnet  has been proposed with the coexistence of antiferroelectricity and altermagnetism  in a single material, which  paves the way for electrically controlled multiferroic devices\cite{k7-3-2}.

\begin{figure*}[t]
    \centering
    \includegraphics[width=0.96\textwidth]{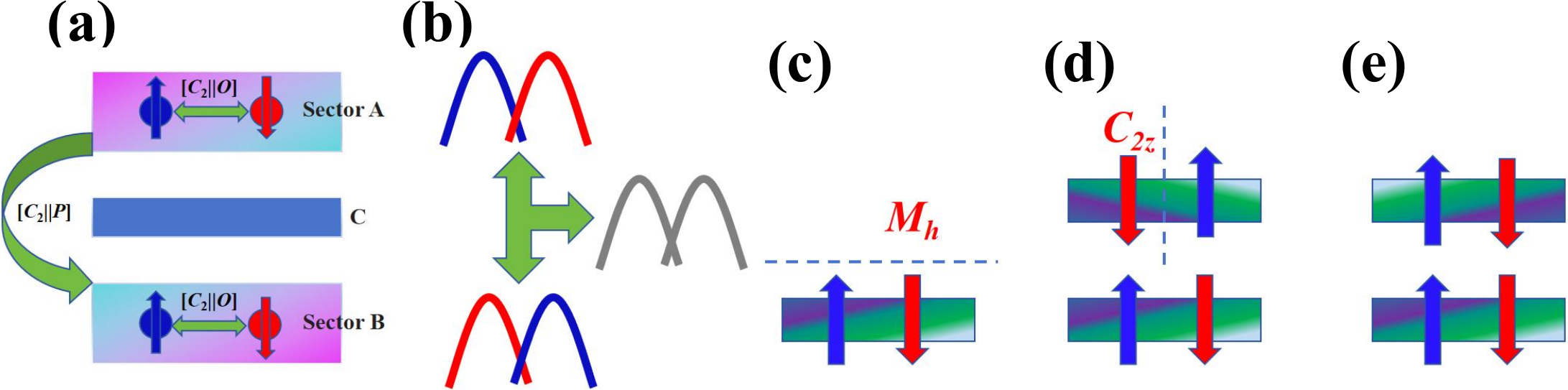}
    \caption{(Color online)(a): For $PT$-symmetrical antiferromagnet, the C plane containing the inversion center separates the
 unit cell into  sector A and B, and  the spin-up and spin-down magnetic atoms of which  are connected
by [$C_2$$\parallel$$O$]   (The $C_2$ is the two-fold rotation perpendicular to the spin axis in the spin space, while $O$ means rotation or mirror operation in the lattice space). The sector A and B are connected by the [$C_2$$\parallel$$P$]. (b): The sector A and B  manifest inverse altermagnetic spin-splitting, which  leads to that the energy bands of $PT$-symmetrical antiferromagnet are at least two-fold degenerate. (c, d, e): Illustration of a general scheme to  build bilayer systems with hidden altermagnetism. (c): A altermagnetic  monolayer  is employed as the fundamental building block, defined as sector B. (d): The upper layer can be obtained from the sector B through a mirror operation with respect to the horizontal dashed line in (c). (e): The sector A is obtained by rotating the upper layer with 180$^{\circ}$ along vertical 2-fold axis in (d), and then the bilayer structure has lattice inversion symmetry $P$. }\label{sy}
\end{figure*}

If two altermagnets  are symmetrically linked by space inversion symmetry ($P$), what interesting physical effects appear in this union  with $PT$ symmetry (the joint symmetry of  $P$ and $T$)?
 In this work, we propose the concept of hidden altermagnetism: the inversion partners consist of two separate altermagnetic sectors, which  produces zero net spin polarization in total, but either of the two inversion-partner sectors possesses  non-zero local spin polarization in the real space.
 Through first-principles calculations, we present
$PT$-symmetric bilayer $\mathrm{Cr_2SO}$ as a example to demonstrat the feasibility of our proposal.
By applying an external electric field along the $z$-direction, the  altermagnetism exhibiting momentum-dependent spin-splitting can be observed. For $PT$-symmetric bilayer $\mathrm{Cr_2SO}$, the intrinsic in-plane magnetization can induce tiny valley polarization, when considering SOC. In fact, a lot of 2D altermagnets can be used as the basic building block to show hidden altermagnetism.
The proposal of hidden altermagnetism considerably broadens the
range of materials for potential antiferromagnetic (AFM) spintronic applications.

\textcolor[rgb]{0.00,0.00,1.00}{\textbf{Concept of hidden altermagnetism.---}}
Even if a magnetic system has $PT$ symmetry, local magnetic atoms of sectors with opposite spins can be connected by rotational or mirror symmetry.
In this study, we introduce the concept of \textit{hidden altermagnetism} (\autoref{sy} (a and b)), which posits that, in a $PT$-symmetric magnetic material, 'local' altermagnetism with momentum-dependent  spin-splitting can arise, when the magnetic atoms with opposite spins within the local environment of specific atomic layer marked with sector A or B are interconnected through [$C_2$$\parallel$$O$]  (The $C_2$/$O$ is the two-fold rotation perpendicular to the spin axis in the spin space/rotation or mirror operation in the lattice space). When two atomic layers (sector A and B) are connected by [$C_2$$\parallel$$P$], their local spin polarization is reversed, so that the overall spin polarization is cancelled to zero. The local spin polarization is essentially a spin-momentum-layer locking effect due to the introduction of the degree of freedom of the  'layer' in the real space.
For energy band structures, the global $PT$ symmetry confirms that: $E_{\uparrow}(k)$=$PT$$E_{\uparrow}(k)$= $P$$E_{\downarrow}(-k)$=$E_{\downarrow}(k)$, resulting in global spin degeneracy or no spin-splitting, while the local [$C_2$$\parallel$$O$] symmetry produces local momentum-dependent  spin-splitting.

It can be difficult to search for bulk or 2D materials with  hidden altermagnetism.
Here, we achieve hidden altermagnetism through an alternate approach of bilayer stacking engineering.
Initially, we take the altermagnetic monolayer as the basic building unit (\autoref{sy} (c)), defined as sector B, to build the bilayer. Through a mirror operation $M_h$ with respect to the horizontal dashed line in \autoref{sy} (c), the upper layer can be derived from the sector B (\autoref{sy} (d)). Subsequently,  the sector A can be obtained by rotating the upper layer in \autoref{sy} (d) with rotation operation $C_{2z}$, producing inversion symmetry $P$=$C_{2z}$$M_h$ in real space for bilayer (\autoref{sy}(e)).
For  2D systems, the wave vector $k$ only has in-plane
components, which leads to spin degeneracy under  [$C_2$$\parallel$$M_h$]  symmetry: $E_{\uparrow}(k)$=[$C_2$$\parallel$$M_h$]$E_{\uparrow}(k)$=$E_{\downarrow}(k)$.
So, the bilayer with [$C_2$$\parallel$$M_h$]  symmetry also possesses hidden altermagnetism.

The fundamental building block with altermagnetism can take any 2D Bravais lattice.
In the bilayer system, the hidden altermagnetism
localized on each layer can  become observable by applying a perpendicular electric field $E$, and the layer-locked unconventional
anomalous magnetic response, such
as the anomalous Halll/Nernst effect  and magneto-optical Kerr effect, can be achieved.
Many 2D altermagnets, such as  $\mathrm{Cr_2O_2}$,  $\mathrm{Cr_2SO}$, $\mathrm{V_2Se_2O}$,  $\mathrm{V_2SeTeO}$ and  $\mathrm{Fe_2Se_2O}$, have been predicted by the first-principles calculations\cite{h13,k60,k6,k7,k7-1,k7-2,k7-3,k7-3-1,k7-3-2}, which  can be used as the basic building unit.
Here, we take Janus $\mathrm{Cr_2SO}$ monolayer as a example to illustrate the concept of hidden altermagnetism.

\begin{figure*}[t]
    \centering
    \includegraphics[width=0.8\textwidth]{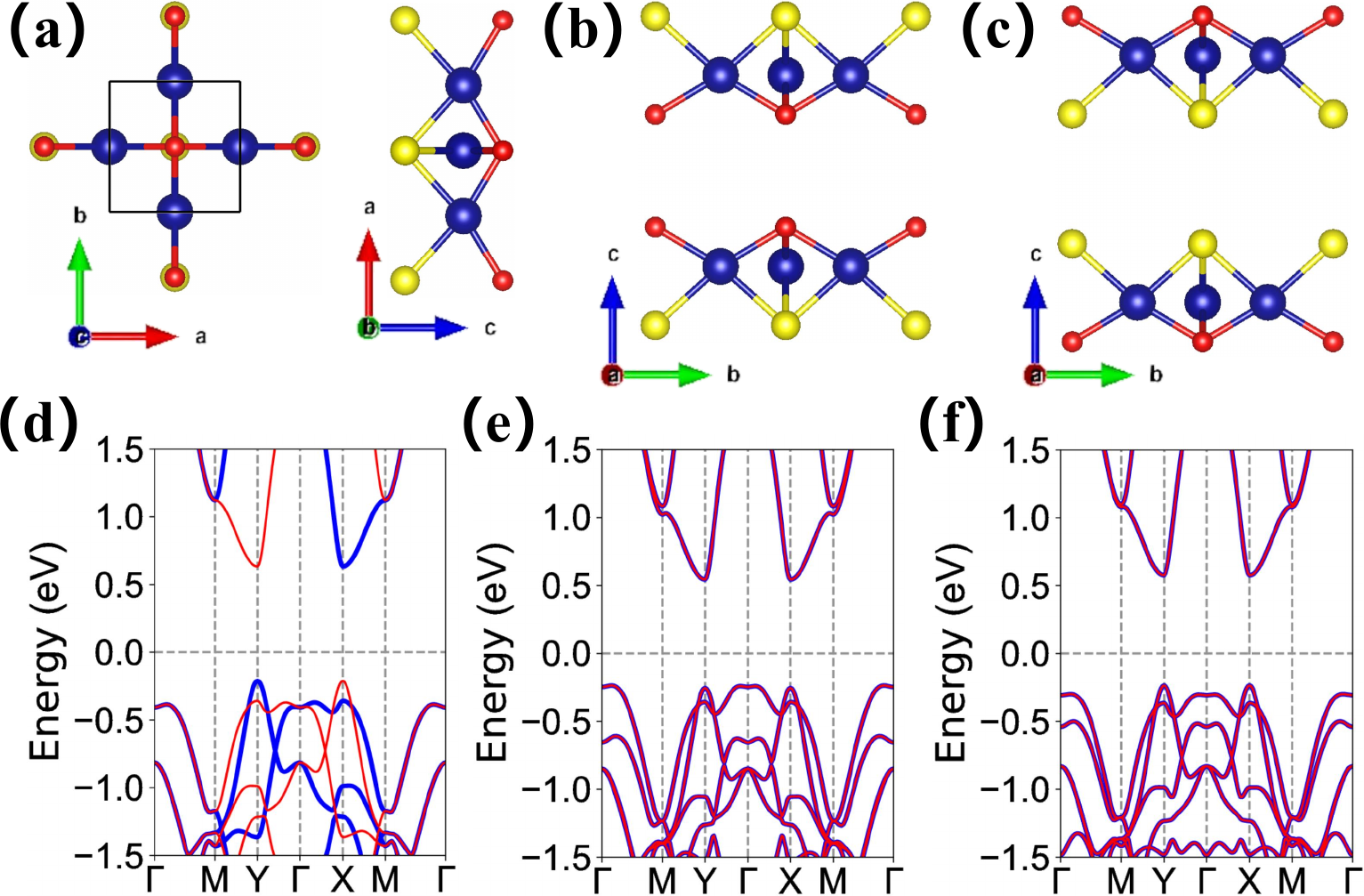}
    \caption{(Color online) For $\mathrm{Cr_2SO}$,  the crystal structure (a, b, c) and  energy band structures (d, e, f) of  monolayer, S-terminal bilayer and O-terminal bilayer. In (a, b, c),  the blue, yellow and red balls  represent Cr, S and
      O atoms, respectively. In  (d, e, f), the spin-up
and spin-down channels are depicted in blue and red.  }\label{str}
\end{figure*}

\textcolor[rgb]{0.00,0.00,1.00}{\textbf{Computational detail.---}}
Density functional theory\cite{1} calculations are carried out
using the Vienna ab initio simulation package (VASP)\cite{pv1,pv2,pv3} by using the projector augmented-wave (PAW) method. The generalized gradient approximation (GGA) of the exchange-correlation functional by
Perdew, Burke, and Ernzerhof (PBE)\cite{pbe} is adopted. The electronic wave
functions are expanded using the plane wave basis set with
a kinetic energy cutoff of 500 eV. Total energy  convergence criterion of  $10^{-8}$ eV and  force convergence criterion of 0.001 $\mathrm{eV.{\AA}^{-1}}$ are set to obtain reliable results.
We add Hubbard correction with $U$=3.55 eV for $d$-orbitals of Cr atoms within the
rotationally invariant approach proposed by Dudarev et al\cite{du}.
The vacuum slab of more than 20 $\mathrm{{\AA}}$ is added to avoid the physical interactions of periodic cells.
A 15$\times$15$\times$1 Monkhorst-Pack $k$-point meshes is used to sample the BZ for structure relaxation and electronic structure calculations. We adopt the dispersion-corrected DFT-D3 method\cite{dft3} to describe the vdW
interactions. The magnetic orientation can be determined by magnetic anisotropy energy (MAE), which can be calculated by $E_{MAE}=E^{||}_{SOC}-E^{\perp}_{SOC}$, in which $||$ and $\perp$  mean that spins lie in
the plane and out-of-plane.

\textcolor[rgb]{0.00,0.00,1.00}{\textbf{Material realization.---}}
Janus monolayer $\mathrm{Cr_2SO}$ with good stability contain three atomic sublayers with two co-planar Cr atoms sandwiched between the O and S atomic layers\cite{k7}, as shown in \autoref{str} (a). Compared to parent monolayer $\mathrm{Cr_2O_2}$ with $P4/mmm$ space group (No.123)\cite{k7-1},  the monolayer  $\mathrm{Cr_2SO}$  crystallizes in the reduced $P4mm$ space group (No.99) due to broken key lattice symmetry $P$. It has been proved that $\mathrm{Cr_2SO}$ possesses altermagnetism with AFM ordering in one unit cell,  and its lattice constants $a$=$b$=3.66 $\mathrm{{\AA}}$. The energy band structures of $\mathrm{Cr_2SO}$ without SOC are plotted in \autoref{str} (d), showing altermagnetic spin splitting of $d$-wave symmetry. The  [$C_2$$\parallel$$M_{xy}$]  symmetry of $\mathrm{Cr_2SO}$ leads to that:$E_{\uparrow}(k_x, k_y)$=[$C_2$$\parallel$$M_{xy}$]$E_{\uparrow}(k_x, k_y)$=$E_{\downarrow}(k_y, k_x)$, giving rise to spin degeneracy along $\Gamma$-M line in BZ. For other  high symmetry paths, the spin-splitting can be observed. It is clearly seen that  two valleys at X and Y high-symmetry points for both conduction and valence bands are  related by [$C_2$$\parallel$$M_{xy}$]  symmetry, producing  spin-valley locking. It is found that $\mathrm{Cr_2SO}$ is  a direct band gap semiconductor with gap value of 0.838 eV, and its valence band maximum (VBM) and conduction band bottom (CBM) are at X/Y point.
The magnetic easy-axis  of $\mathrm{Cr_2SO}$ is in-plane  with MAE being  -93 $\mathrm{\mu eV}$/Cr.
When including SOC, the valley polarization between X and Y valleys of $\mathrm{Cr_2SO}$ with in-plane  magnetization along $x$ or $y$ direction  will arise due to broken $M_{xy}T$ symmetry\cite{k60}.

\begin{figure*}[t]
    \centering
    \includegraphics[width=0.96\textwidth]{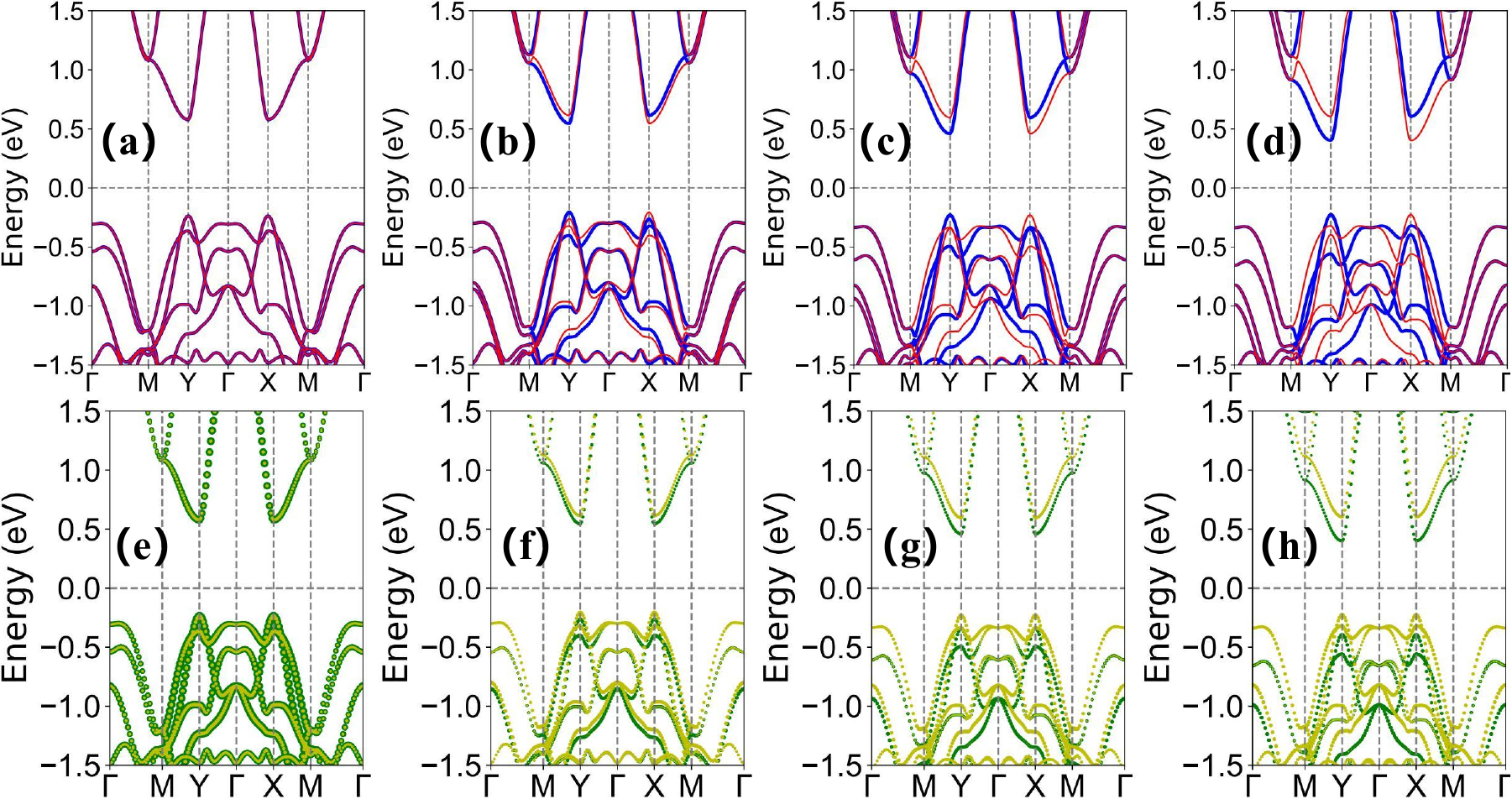}
     \caption{(Color online)For O-terminal bilayer  $\mathrm{Cr_2SO}$, the energy band structures (a, b, c, d) with layer-characteristic projection (e, f, g, h) at  representative $E$=+0.00, +0.01, +0.02 and +0.03  $\mathrm{V/{\AA}}$.
     In (a, b, c, d),  the blue (red) represents spin-up (spin-down) characters. In (e, f, g, h),
      the yellow (green) represents  lower (upper)-layer characters. }\label{band}
\end{figure*}
\begin{figure}[t]
    \centering
    \includegraphics[width=0.45\textwidth]{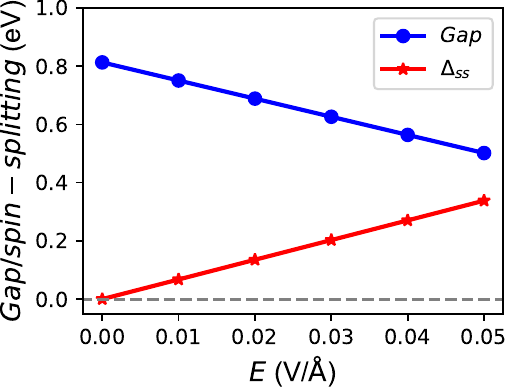}
     \caption{(Color online) For O-terminal bilayer  $\mathrm{Cr_2SO}$, the energy band gap ($Gap$) and the spin-splitting ($\Delta_{ss}$)  between the first and second conduction bands at X/Y point as a function of electric field $E$.}\label{gap}
\end{figure}

\begin{figure}[t]
    \centering
    \includegraphics[width=0.45\textwidth]{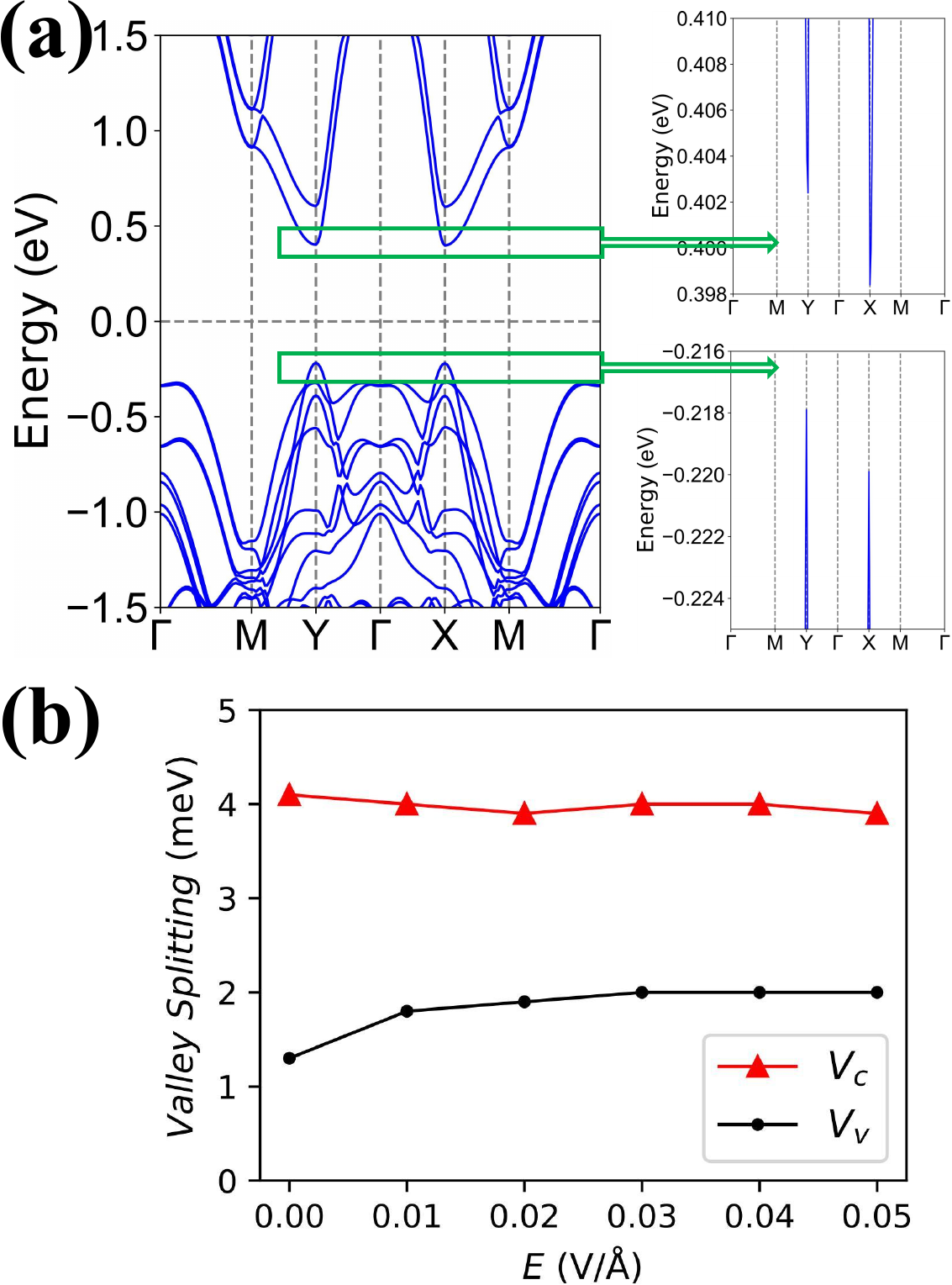}
     \caption{(Color online) For O-terminal bilayer  $\mathrm{Cr_2SO}$, (a): the energy band structures  with SOC along with the  enlarged views near the Fermi energy level at  representative $E$=+0.03  $\mathrm{V/{\AA}}$. (b): the valley splitting between X and Y valleys for both conduction ($V_C$) and valence ($V_V$) bands as a function of electric field $E$.}\label{band-1}
\end{figure}

Next, we build  $PT$-symmetric bilayers using the design procedure presented in \autoref{sy}.
Because $\mathrm{Cr_2SO}$ has $C_2$ rotation symmetry, we can simply do the operations shown in \autoref{sy} (c) and (d), producing S-terminal bilayer and O-terminal bilayers with $P$ lattice symmetry (\autoref{str} (b) and (c)).
To determine the ground state of these bilayers, the intralayer AFM and interlayer AFM  (AFM1), and intralayer AFM and interlayer ferromagnetic (FM)  (AFM2) configurations are considered (See FIG.S1\cite{bc}). Calculated results show that  S-terminal/O-terminal bilayer possesses AFM1 ordering, which is 1.1/3.0 meV per unit cell lower than that of AFM2 case. The optimized lattice constants are $a$=$b$=3.645/3.645 $\mathrm{{\AA}}$ by GGA+$U$ for  S-terminal/O-terminal bilayer with AFM1 case, which is slightly smaller than that of monolayer. The AFM1 ordering possesses $PT$ symmetry, which can produce hidden altermagnetism. The energy band structures  of  S-terminal bilayer and O-terminal bilayer are plotted in \autoref{str} (e) and (f) without SOC. The S-terminal bilayer is an indirect bandgap semiconductor of 0.78 eV with VBM at one point along $\Gamma$-M line and CBM at X/Y point. However, the O-terminal bilayer is still a direct bandgap semiconductor of 0.81 eV with VBM/CBM at X or Y point.

 In some regions of the BZ (for example: around $\Gamma$), the two layers are coupled strongly, which results in that the energy spectrum of bilayer is different from that of monolayer. The interlayer coupling of  electronic states along some wave vector directions (For example: Y-M and X-M paths) in the BZ is effectively restrained.
If the bands of  bilayer represent a straightforward superposition of  bands of two monolayers, then the energy bands along the $\Gamma$-M path should exhibit quadruple degeneracy. However, the first two bands in the valence band along the $\Gamma$-M path are doubly degenerate, which means that the interlayer coupling of  electronic states is strong.
These indicate that our proposed $PT$-symmetric bilayer is not a simple overlap of two monolayer altermagnetic  $\mathrm{Cr_2SO}$.

If we want to lift the degeneracy of $PT$-symmetric magnetic material, either $P$ or $T$ symmetry
should be broken.  This $P$-symmetry breaking can be achieved by applying an external electric field along the $z$-direction.
Here, we use O-terminal bilayer $\mathrm{Cr_2SO}$ as an example to illustrate the effect of the electric field on spin-splitting in hidden altermagnetism.  For S-terminal bilayer case, the similar results can be obtained.
Firstly,   we determine the magnetic ground state under $+z$ electric field  (0.00-0.05 $\mathrm{V/{\AA}}$)  by
the energy difference between  AFM2 and AFM1 configurations.
Within considered $E$ range,   the AFM1 ordering is always ground state from FIG.S2\cite{bc}.

The energy band structures along  with cases of  layer-characteristic projection  at  representative $E$=+0.00, +0.01, +0.02 and +0.03  $\mathrm{V/{\AA}}$ without SOC are plotted in \autoref{band}.   When electric field is applied, it is clearly seen that there is altermagnetic spin-splitting due to layer-dependent electrostatic potential
caused by  electric field\cite{k9,k10}.  Based on  layer-characteristic projection, the real-space segregation of spin polarization can be observed, which make hidden altermagnetism to be observed experimentally. The bilayer still has [$C_2$$\parallel$$M_{xy}$]  symmetry with out-of-plane electric field, which leads to spin degeneracy along $\Gamma$-M line in BZ. Therefore, by applied electric field, the first two bands in the valence band along the $\Gamma$-M path are still doubly spin-degenerate, but the third bands change from a quadruple degeneracy to two double spin-degeneracy. In the absence of an electric field, the first two bands in the valence band along the $\Gamma$-M path constitute a mixture of  the upper- and lower-layer characters. With increasing $E$, the first  band is dominated by lower-layer, and the second   band is from upper-layer character.

The energy band gap  and the spin-splitting  between the first and second conduction bands at X/Y point as a function of electric field $E$ are shown in \autoref{gap}.  It is clearly seen that both gap and spin-splitting vs $E$ show a linear relationship, and the gap/spin-splitting decreases/increases with increasing $E$.
In fact, the spin-splitting  can be approximately calculated by  $eEd$\cite{k10}, where $e$ and $d$ denote the electron charge and the interlayer distance of two Cr layers. Taking $E$=+0.03$\mathrm{V/{\AA}}$ as a example with the $d$ being  6.91 $\mathrm{{\AA}}$,  the estimated spin-splitting  is approximately 207 meV, being very close to the first-principle result of 203 meV.
When the direction of electric field is reversed, the order of spin-splitting/layer-character is reversed (see FIG.S3\cite{bc}).

Although the spin-splitting of altermagnetism does not require the help of SOC, we consider the effect of SOC on the energy band structures of O-terminal bilayer $\mathrm{Cr_2SO}$.
With SOC, the magnetization direction can produce  important influences on electronic structures of tetragonal magnetic materials by changing  magnetic group symmetry\cite{k60}.  The MAE of O-terminal bilayer $\mathrm{Cr_2SO}$ as a function of $E$ is plotted in FIG.S4\cite{bc}, and  the negative MAE confirms that its easy axis  is in-plane $x$/$y$ direction within considered $E$ range.
The valley polarization will arise when the orientation of magnetization, for example in-plane $x$/$y$ direction, breaks the $C_{4z}T$ or $M_{xy}T$ symmetry. When the magnetization direction switches between the $x$ and $y$ direction, the valley polarization will be reversed. The energy band structures of O-terminal bilayer $\mathrm{Cr_2SO}$ with SOC for in-plane $x$ magnetization direction  at  representative $E$= +0.03  $\mathrm{V/{\AA}}$ are plotted in \autoref{band-1} (a). The valley polarization can be observed between $Y$ and $X$ valleys, and the valley-splitting ( $\Delta E_{VS}=E_{Y}-E_{X}$)  in the conduction/Valence bands is about 4.0/2.0 meV.
The valley-splitting between Y and X valleys for both conduction  and valence  bands as a function of $E$ are shown in \autoref{band-1} (b). Calculated results show that the increased $E$ has a small effect on valley-splitting.

\textcolor[rgb]{0.00,0.00,1.00}{\textbf{Discussion and Conclusion.---}}
In general, in crystals with inversion-asymmetric sectors,  HSP can not be directly measured   without breaking $PT$ symmetry. However, by using spin- and angle-resolved photoemission spectroscopy
(ARPES) measurements, the HSP effect has been experimentally confirmed in many  materials\cite{h7,h8,h9,h10,h11}.
 Therefore, our proposed hidden altermagnetism can be confirmed in experiment. Compared to conventional antiferromagnets,
altermagnets have demonstrated a series of phenomena, including anomalous Hall/Nernst effect, nonrelativistic spin-polarized currents and the magneto-optical Kerr effect\cite{h13}.  By adding the degree of freedom of the "layer" in real space for  hidden altermagnetism, the layer-locked  phenomena possessed by altermagnets may be realized.
Supreme to ferromagnets, antiferromagnet
 exhibits tremendous potential for spintronic devices with high immunity to magnetic field disturbance
thanks to their intrinsic advantages of zero stray field and terahertz dynamics\cite{z1,z2}. Hidden altermagnet  constitutes a special class within antiferromagnet, and can add more material basis and exotic physical insights to the development of spintronics.

 In summary, we report the possible concept of hidden altermagnetism, which possess an hidden altermagnetic  spin-splitting.
The hidden altermagnetism requires that the system possesses high
global $PT$ symmetry, but is comprised of individual sectors with  local  altermagnetic ordering. Taking the $PT$-symmetric bilayer $\mathrm{Cr_2SO}$  as
a representative, we demonstrate that the hidden altermagnetism can be achieved, and  an  out-of-plane external electric field can be used to separate and detect the hidden altermagnetism  in experiment.
Our findings thus
open new perspectives for the HSP research, and advance relevant theories and experiments to search for real material with hidden altermagnetism, and then to explore the intriguing
physics of altermagnetic HSP.

\begin{acknowledgments}
This work is supported by Natural Science Basis Research Plan in Shaanxi Province of China  (2021JM-456). We are grateful to Shanxi Supercomputing Center of China, and the calculations were performed on TianHe-2. We thank Prof. Guangzhao Wang for providing VASP software and helpful discussions.
\end{acknowledgments}

\end{document}